\documentclass[%
prl,
 reprint, 
 amsmath,amssymb,
 aps,preprintnumbers]{revtex4-1}

\usepackage[colorlinks=true,linkcolor=black, citecolor=black,
urlcolor=black]{hyperref}

\usepackage{multirow,graphics}
    \newcommand{\beq}{\begin{equation}}
    \newcommand{\eeq}{\end{equation}}
    \newcommand\beqa{\begin{eqnarray}}
    \newcommand\eeqa{\end{eqnarray}}

\usepackage{amstext}
\usepackage{amssymb}
\usepackage{amsmath}
\usepackage{graphicx}
\usepackage{color}

\newcommand{\dhd}{{\textstyle d}
\lower.03ex\hbox{\kern-0.38em$^{\scriptstyle-}$}\kern-0.05em{}}
\newcommand{\dbar}{{\textstyle \delta}
\lower.03ex\hbox{\kern-0.38em$^{\scriptstyle-}$}\kern-0.05em{}}
\newcommand{\half}{{1\over 2}}

\newcommand{\balfa}{{\bar \alpha}}

\newcommand{\bsi}{{\bar \psi}}

\newcommand{\cald}{{\cal D}}  
\newcommand{\calf}{{\cal F}}

\newcommand{\calo}{{\cal O}}

\newcommand{\ticalo}{\tilde {\cal O}}

\newcommand{\tigma}{\tilde {\sigma}}

\begin{document}

\newcommand{\IM}{{\rm Im}\,}
\newcommand{\card}{\#}
\newcommand{\la}[1]{\label{#1}}
\newcommand{\eq}[1]{(\ref{#1})}
\newcommand{\figref}[1]{Fig. \ref{#1}}
\newcommand{\abs}[1]{\left|#1\right|}
\newcommand{\comD}[1]{{\color{red}#1\color{black}}}

\makeatletter
     \@ifundefined{usebibtex}{\newcommand{\ifbibtexelse}[2]{#2}} {\newcommand{\ifbibtexelse}[2]{#1}}
\makeatother

\preprint{JLAB-THY-19-2940}

\newcommand{\footnoteab}[2]{\ifbibtexelse{%
\footnotetext{#1}%
\footnotetext{#2}%
\cite{Note1,Note2}%
}{%
\newcommand{\textfootnotea}{#1}%
\newcommand{\textfootnoteab}{#2}%
\cite{thefootnotea,thefootnoteab}}}

\def\e{\epsilon}
     \def\bT{{\bf T}}
    \def\bQ{{\bf Q}}
    \def\wT{{\mathbb{T}}}
    \def\wQ{{\mathbb{Q}}}
    \def\ttQ{{\bar Q}}
    \def\tQ{{\tilde \bP}}
        \def\bP{{\bf P}}
    \def\CF{{\cal F}}
    \def\cC{\CF}
     \def\Tr{\text{Tr}}
     \def\l{\lambda}
\def\hbZ{{\widehat{ Z}}}
\def\bZ{{\resizebox{0.28cm}{0.33cm}{$\hspace{0.03cm}\check {\hspace{-0.03cm}\resizebox{0.14cm}{0.18cm}{$Z$}}$}}}
\newcommand{\rb}{\right)}
\newcommand{\lb}{\left(}

\newcommand{\gT}{T}\newcommand{\gQ}{Q}

\title{Conformal invariance of TMD rapidity evolution }

\author{ Ian Balitsky$^{a}$ and Giovanni A. Chirilli$^{b}$}

\affiliation{%
\(^{a}\)Physics Dept., Old Dominion University, Norfolk VA 23529 \&
 Theory Group, JLAB, 12000 Jefferson Ave, Newport News, VA 23606
\\
\(^{b}\)Institut f\"ur Theoretische Physik, Universit\"at Regensburg,\\ D-93040 Regensburg, Germany\\
}

\begin{abstract}
We discuss  conformal properties of TMD operators and present the 
result of the conformal rapidity evolution of TMD operators in the Sudakov region.
 \end{abstract}

 \maketitle

\section{Introduction}
In recent years, the  transverse-momentum dependent parton distributions (TMDs) 
\cite{Collins:1981uw,Collins:1984kg,Ji:2004wu,GarciaEchevarria:2011rb}  have been widely used in the analysis of 
 processes like semi-inclusive deep inelastic scattering 
or particle production in hadron-hadron collisions (for a review, see Ref. \cite{Collins:2011zzd}). 

The TMDs are defined as matrix elements of quark or gluon operators with attached light-like gauge links
(Wilson lines) going to either $+\infty$ or $-\infty$ depending on the process under consideration. 
It is well known that these TMD operators exhibit rapidity divergencies due to infinite light-like gauge links 
and the corresponding rapidity/UV divergences  should be regularized. There are two schemes on the market: 
the most popular is based on CSS  \cite{Collins:1984kg} or SCET \cite{Rothstein:2016bsq} formalism and the second one is 
adopted from the small-$x$ physics \cite{Lipatov:1996ts,Kovchegov:2012mbw}.
The obtained evolution equations differ even at the leading-order level and
 need to be reconciled, especially in view of the future EIC accelerator which will probe the 
TMDs at values of Bjorken $x$ between small-$x$ and $x\sim 1$ regions. 

In our opinion, a good starting point is to obtain conformal leading-order evolution equations. 
It is well known that  at the leading order pQCD is conformally invariant so there is a hope to get 
any evolution equation without explicit running coupling from conformal considerations. 
In our case, since TMD operators are defined with attached light-like Wilson lines,
formally they will transform covariantly under the subgroup of full conformal group 
which preserves this light-like direction. 
However, as we mentioned, the TMD operators contain rapidity divergencies which need to be regularized. 
At present, there is no rapidity cutoff which preserves conformal invariance so the best one can do is to find
the cutoff which is conformal at the leading order in perturbation theory. In higher orders, one should not 
expect conformal invariance since it is broken by running of QCD coupling. However, if one considers corresponding 
correlation functions in  ${\cal N}=4$ SYM, one should expect conformal invariance. 
After that, the results obtained in ${\cal N}=4$ SYM theory can be used as a starting point of QCD calculation. Typically, 
the result in ${\cal N}=4$ theory gives the most complicated part of pQCD result, i.e. the one with maximal transcendentality.
Thus, the idea is to find TMD operator conformal in ${\cal N}=4$ SYM and use it in QCD. 
 This scheme was successfully applied to the
 rapidity evolution of color dipoles. At the leading order, the BK evolution of color dipoles 
 \cite{Balitsky:1995ub,Balitsky:1998ya,Kovchegov:1999yj,Kovchegov:1999ua} is invariant under SL(2,C) 
 group. At the NLO order the ``conformal dipole'' with $\alpha_s$ correction \cite{Balitsky:2009xg} makes NLO BK evolution Mobius
 invariant for ${\cal N}=4$ SYM and the corresponding QCD kernel \cite{Balitsky:2008zza} differs by  terms proportional to $\beta$-function.

\section{Conformal invariance of TMD operators}

For definiteness, we will talk first about gluon operators with light-like Wilson lines stretching to $-\infty$ in ``+'' direction.
The gluon TMD (unintegrated gluon distribution) is defined as \cite{Mulders:2000sh}
\begin{eqnarray}
&&\hspace{-0mm}
\cald(x_B,k_\perp,\eta)~=~\!\int\!d^2z_\perp~e^{i(k,z)_\perp}\cald(x_B,z_\perp,\eta),
\label{TMDg}\\
&&\hspace{-0mm}
g^2\cald(x_B,z_\perp,\eta)~=~{-x_B^{-1}\over 2\pi p^-}\!\int\! dz^+ ~e^{-ix_Bp^-z^+}
\nonumber\\
&&\hspace{-0mm}
\times~ \langle P|\calf^a_\xi(z)
[z-\infty n,-\infty n]^{ab}\calf^{b\xi}(0)|P\rangle\Big|_{z^-=0}
\nonumber
\end{eqnarray}
where $|P\rangle$ is an unpolarized target with momentum $p\simeq p^-$ (typically proton) and $n=({1\over\sqrt{2}},0,0,{1\over\sqrt{2}})$ is a light-like vector 
in ``+'' direction.
Hereafter we use the notation
\begin{equation}
\hspace{-0mm}
\calf^{\xi,a}(z_\perp,z^+)~\equiv~gF^{- \xi,m}(z)[z,z-\infty n]^{ma}\Big|_{z^-=0}
\label{kalf}
\end{equation}
where $[x,y]$ denotes straight-line gauge link connecting points $x$ and $y$:
\begin{equation}
~[x,y]~\equiv~{\rm P}e^{ig\int\! du~(x-y)^\mu A_\mu(ux+(1-u)y)}
\label{defu}
\end{equation}
To simplify  one-loop evolution
we multiplied $F_{\mu\nu}$ by coupling constant. Since the $gA_\mu$ is renorm-invariant we do not need to consider self-energy diagrams (in 
the background-Feynman gauge). Note that $z^-=0$ is fixed  by
the original factorization formula for particle production  \cite{Collins:2011zzd} (see also the discussion in 
Ref. \cite{Balitsky:2017flc,Balitsky:2017gis}).

The algebra of full conformal group $SO(2,4)$  consists of four operators $P^\mu$, six $M^{\mu\nu}$, four 
special conformal generators $K^\mu$, and dilatation operator $D$.
It is easy to check that in the leading order the following 11 operators act on gluon TMDs covariantly
\begin{equation}
P^i,P^-,M^{12},M^{-i},D,K^i,K^-,M^{-+}
\label{generators}
\end{equation}
while the action of operators $P^+,M^{+i}$, and $K^+$ do not preserve the  form of the operator (\ref{kalf}).
The action of the generators (\ref{generators}) on the operator (\ref{kalf}) is the same as the action on the field $F^{-i}$
without gauge link attachments. 
The corresponding group consists of transformations which leave the hyperplane $z^-=0$
and vector $n$ 
invariant. Those include 
shifts in transverse and $``+''$ directions, rotations in the 
transverse plane, Lorentz rotations/boosts created by $M^{-i}$, dilatations, and special conformal
transformations
\begin{equation}
z'_\mu~=~{z_\mu -a_\mu z^2\over 1-2a\cdot z+a^2z^2}
\label{speconf}
\end{equation}
with $a=(a^+,0,a_\perp)$. In terms of ``embedding formalism'' \cite{Dirac:1936fq,Mack:1969rr,Ferrara:1973yt,Rychkov:2016iqz} defined in  6-dim space, 
 this subgroup is isomorphic 
to  ``Poincare + dilatations'' group of the 4-dim  subspace orthogonal to our physical light-like ``+'' and ``-'' directions.

As we noted, infinite Wilson lines in the definition (\ref{kalf}) of TMD operators make 
them divergent. As we discussed above, it is very advantageous to have a cutoff 
of these divergencies compatible with approximate conformal invariance of tree-level QCD.
The evolution equation with such cutoff should be invariant with respect to transformations 
described above.

In the next Section we demonstrate that the ``small-x'' rapidity cutoff enables us to
get a conformally invariant evolution of TMD in the so-called Sudakov region. 

\section{TMD factorization in the Sudakov region}

The rapidity evolution of TMD operator (\ref{TMDg})
is very different in the region of large and small longitudinal separations $z^+$. 
The evolution at small $z^+$ is linear and double-logarithmic
while  at large $z^+$ the evolution become non-linear
due to the production of color dipoles typical for small-$x$ evolution. 
It is convenient to consider as a starting point the simple case of
TMD evolution in the so-called Sudakov region corresponding to
small longitudinal distances. 

First,  let us specify what we  call a Sudakov region. 
A typical  
factorization formula for the differential cross section  of particle production 
in hadron-hadron collision is \cite{Collins:2011zzd, Collins:2014jpa}
\begin{eqnarray}
&&\hspace{-2mm}
{d\sigma\over  d\eta d^2q_\perp}~=~
\sum_f\!\int\! d^2b_\perp e^{i(q,b)_\perp}
\cald_{f/A}(x_A,b_\perp,\eta)
\nonumber\\
&&\hspace{-2mm}
\times~
\cald_{f/B}(x_B,b_\perp,\eta)\sigma(ff\rightarrow H)~+~...
\label{TMDf}
\end{eqnarray}
where $\eta=\half\ln{q^+\over q^-}$ is the rapidity, $\cald_{f/h}(x,z_\perp,\eta)$ is the 
TMD density of  a parton $f$  in hadron $h$, and $\sigma(ff\rightarrow H)$ is the cross section of production of particle $H$ 
of invariant mass $m_H^2=q^2\equiv Q^2$ in the scattering of two partons.  
(One can keep in mind Higgs production in the approximation of point-like gluon-gluon-Higgs vertex).  
The Sudakov region is defined by  $Q\gg q_\perp\gg 1$GeV since at such kinematics 
there is a double-log evolution for transverse momenta between $Q$ and $q_\perp$. In the coordinate space,
TMD factorization  (\ref{TMDf})  looks like
\begin{eqnarray}
&&\hspace{-0mm}
\langle p_A,p_B| F^a_{\mu\nu}F^{a\mu\nu}(z_1)  F^b_{\lambda\rho}F^{b\lambda\rho}(z_2)|p_A,p_B\rangle
\nonumber\\
&&\hspace{-1mm}
=~{1\over N_c^2-1}\langle p_A|\ticalo_{ij}(z_1^-,z_{1_\perp};z_2^-,z_{2_\perp}) |p_A\rangle^{\sigma_A}
\nonumber\\
&&\hspace{-1mm}
\times \langle p_B|\calo^{ij}(z_1^+,z_{1_\perp};z_2^+,z_{2_\perp}) |p_B\rangle^{\sigma_B}~+...
\label{facoord}
\end{eqnarray}
where
\begin{eqnarray}
&&\hspace{-2mm}
\calo_{ij}(z_1^+,z_{1_\perp};z_2^+,z_{2_\perp})
\nonumber\\
&&\hspace{-2mm}=~\calf^a_i(z_1)
[z_1-\infty n,z_2-\infty n]^{ab}\calf^b_j(z_2)\Big|_{z_1^-=z_2^-=0}\,,
\label{kalo}
\\
&&\hspace{-2mm}
\ticalo_{ij}(z_1^-,z_{1_\perp};z_2^-,z_{2_\perp})
\nonumber\\
&&\hspace{-2mm}
=~
\calf^a_i(z_1)
[z_1-\infty n',z_2-\infty n']^{ab}\calf^b_j(z_2)\Big|_{z_1^+=z_2^+=0}\,,
\nonumber\\
&&\hspace{-2mm}
\calf^{i,a}(z_\perp,z^-)~\equiv~F^{+i,m}(z)[z,z-\infty n']^{ma}\Big|_{z^+=0}\,.
\label{tikalo}
\end{eqnarray}
Here 
$p_A=\sqrt{s\over 2}n+{p_A^2\over\sqrt{2s}}n',~p_B=\sqrt{2\over s}n'+{p_B^2\over\sqrt{2s}}n$
and $n'=\big({1\over\sqrt{2}},0,0,-{1\over\sqrt{2}}\big)$. Our metric is $x^2=2x^+x^- - x_\perp^2$.

As we mentioned, TMD operators exhibit rapidity divergencies due to infinite light-like gauge links.
The ``small-$x$ style'' rapidity cutoff for longitudinal divergencies is imposed
as the upper limit of $k^+$ components of gluons emitted from the 
Wilson lines. As we will see below, to get the conformal invariance of the leading-order evolution
we need to impose the cutoff of $k^+$ components of gluons correlated with transverse size of TMD in the following way:
\begin{eqnarray}
&&\hspace{-0mm} 
\big(\calf^{i,a}(z_\perp,z^+)\big)^\sigma
~\equiv~F^{- i,m}(z)\big[{\rm P}e^{ig\!\int_{-\infty}^{z^+}\!\! dz^+ A^{-,\sigma}(up_1+x_\perp)}\big]^{ma},
\nonumber\\
&&\hspace{-0mm} 
A^\sigma_\mu(x)~=~\int\!{d^4 k\over 16\pi^4} ~\theta\Big({\sigma\sqrt{2}\over z_{12_\perp}}-|k^+|\Big)e^{-ik\cdot x} A_\mu(k)
\label{cutoff}
\end{eqnarray}
Similarly, the operator $\ticalo$ in Eq. (\ref{tikalo}) is defined with 
with the rapidity cutoff for $\beta$ integration imposed as $\theta\big({\tigma\sqrt{2}\over z_{12_\perp}}-|k^-|\big)$.

The Sudakov region $Q^2\gg q_\perp^2$ in the coordinate space corresponds to
\begin{equation}
z_{12_\parallel}^2~\equiv~2z_{12}^-z_{12}^+~\ll~z_{12_\perp}^2
\label{sudkoord}
\end{equation}
In the leading log approximation, 
the upper cutoff for $k^+$ integration in the target matrix element in Eq. (\ref{facoord}) is 
$\sigma_B={1\over \sqrt{2}}{z_{12_\perp}\over z^-_{12}}$ and  similarly the $\beta$-integration 
cutoff in projectile matrix element is $\sigma_A={1\over \sqrt{2}}{z_{12_\perp}\over z^+_{12}}$. 

In the next Section we demonstrate that rapidity cutoff (\ref{cutoff}) enables us to
get a conformally invariant evolution of TMD in the Sudakov region (\ref{sudkoord}). 

\section{One-loop evolution of TMDs}
\subsection{Evolution of gluon TMD operators in the Sudakov region}
In this Section we derive the evolution of gluon TMD operator (\ref{kalo}) with respect to cutoff $\sigma$
in the leading log approximation.
%
\begin{figure}[htb]
\begin{center}
\includegraphics[width=55mm]{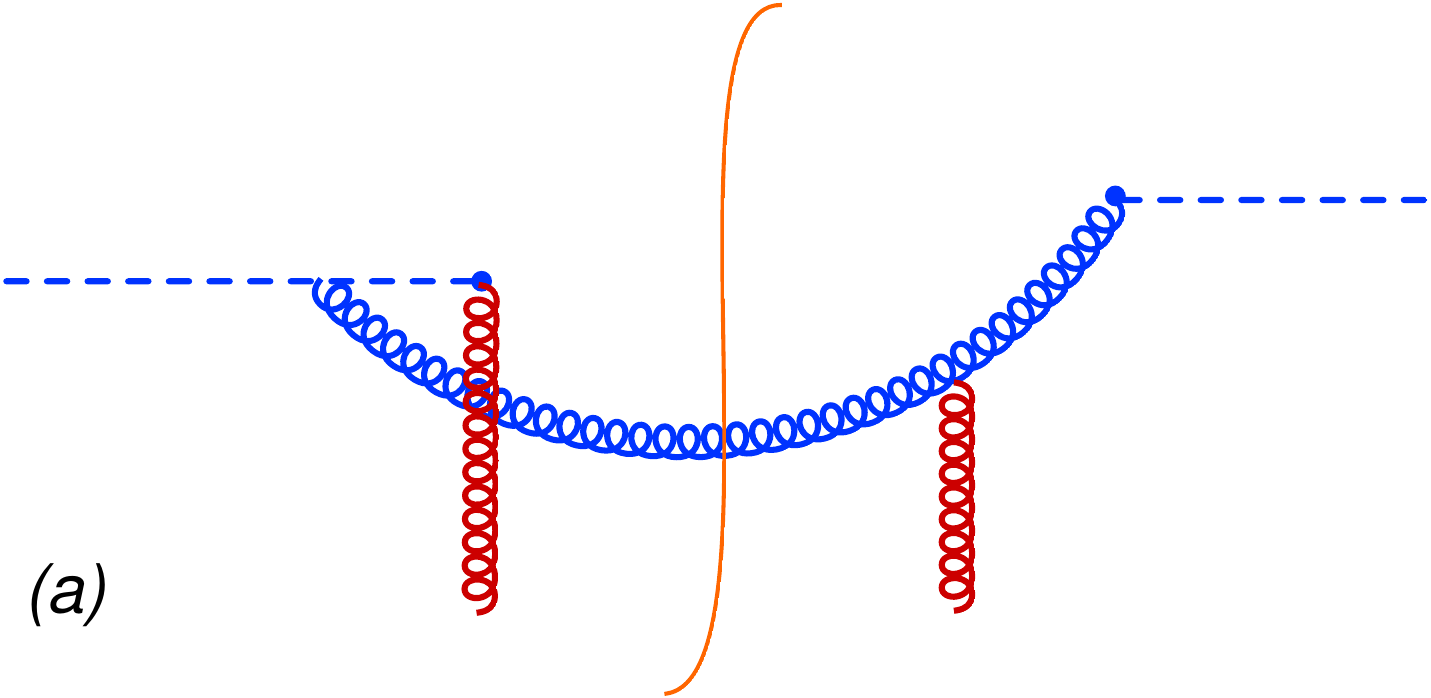}
\hspace{12mm}
\includegraphics[width=55mm]{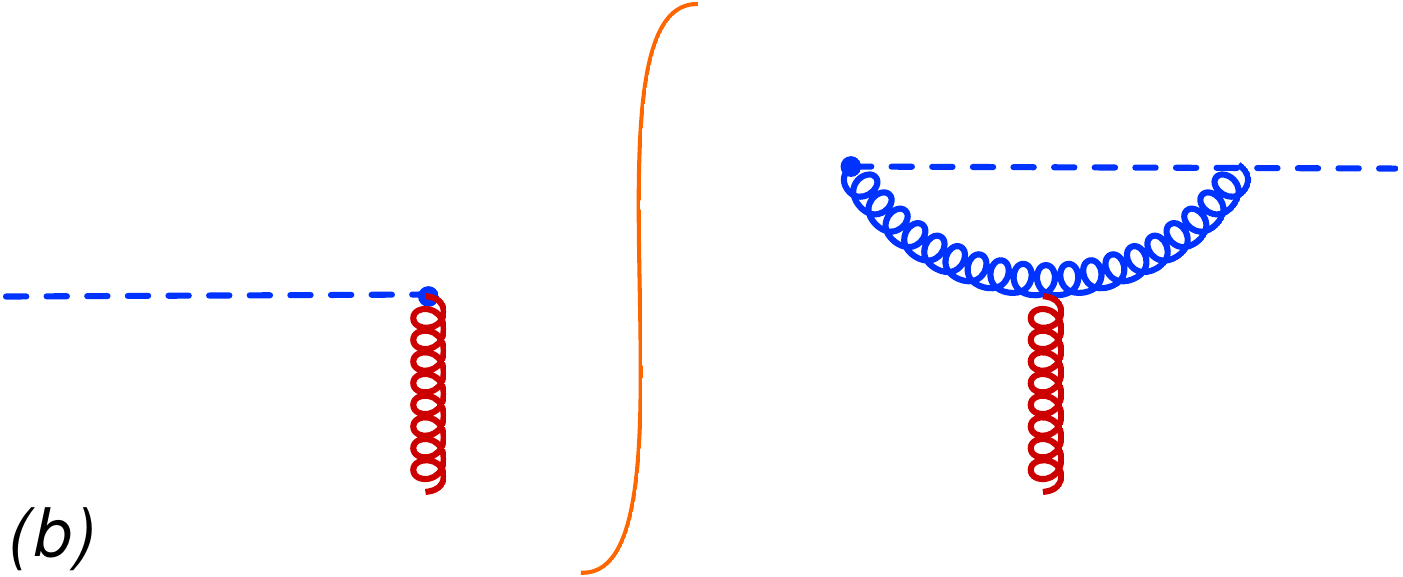}
\end{center}
\caption{Typical diagrams for production (a) and virtual (b) contributions to the evolution kernel. The dashed lines denote gauge links.\label{fig:2}}
\end{figure}
%
As usual, to get an evolution equation we integrate over momenta ${\sigma_2\sqrt{2}\over z_{12_\perp}}>k^+>{\sigma_1\sqrt{2}\over z_{12_\perp}}$. 
To this end, we calculate diagrams shown in Fig. 1 in the background field of gluons with $k^+<{\sigma_1\sqrt{2}\over z_{12_\perp}}$.
The calculation is easily done by method developed in Refs. \cite{Balitsky:2015qba,Balitsky:2016dgz} 
and the result is
\begin{equation}
\calo^{\sigma_2}(z_1^+,z_2^+)
~=~{\alpha_sN_c\over 2\pi}\!\int\limits_{\sigma_1\sqrt{2}\over |z_{12_\perp}|}^{\sigma_2\sqrt{2}\over |z_{12_\perp}|}\!
{dk^+\over k^+}~K\calo^{\sigma_1}(z_1^+,z_2^+)
\label{sudeqn1}
\end{equation}
where the kernel $K$ is given by
\begin{eqnarray}
&&\hspace{-1mm}
K\calo(z_1^+,z_2^+)~
\label{kernel}\\
&&\hspace{-1mm}
=~\calo(z_1^+,z_2^+)
\!\int_{-\infty}^{z_1^+}\! {dz'^+\over z_2^+-z'^+}e^{-i{|z_{12_\perp}|\sigma\over \sqrt{2}(z_2-z')^+}}
\nonumber\\
&&\hspace{-1mm}
+~\calo(z_1^+,z_2^+)
\!\int_{-\infty}^{z_2^+}\! {dz'^+\over z_1^+-z'^+}e^{i{|z_{12_\perp}|\sigma\over \sqrt{2}(z_1-z')^+}}
\nonumber\\
&&\hspace{-1mm}
~-\!\int^{z_1^+}_{-\infty} \! dz'^+ 
 {\calo(z_1^+,z_2^+)-\calo(z'^+_1,z_2^+)\over z_1^+ -z'^+}
\nonumber\\
&&\hspace{-1mm}
~-\!\int^{z_2^+}_{-\infty} \! dz'^+ 
 {\calo(z_1^+,z_2^+)-\calo(z_1^+,z'^+_2)\over z_2^+-z'^+ }
 \nonumber
\end{eqnarray}
where we suppress arguments $z_{1_\perp}$ and $z_{2_\perp}$ since they do not change 
during the evolution in the Sudakov regime.
The first two terms in the kernel $K$ come from the ``production'' diagram in Fig. 1a while the last two terms from 
``virtual'' diagram in Fig. 1b. 
The result (\ref{kernel}) can be also obtained from Ref. \cite{Balitsky:2016dgz}  by Fourier transformation of Eq. (5.9) 
with the help of Eqs. (3.12) and (3.30) therein. The approximations  for diagrams in Fig. 1 leading to Eq. (\ref{kernel}) are valid as long as 
\begin{equation}
k^+\gg {z^+_{12}\over z_{12_\perp}^2}
\label{sudapprox}
\end{equation}
which gives the region of applicability of Sudakov-type evolution.

Evolution equation (\ref{sudeqn1}) can be easily integrated using Fourier transformation. 
Since 
\begin{eqnarray}
&&\hspace{-0mm}
Ke^{-ik^- z_1^+ + ik '^-z_2^+}~
=~\Big[
-2\ln \sigma z_{12_\perp}-\ln(ik^-)-\ln(-ik'^-)
\nonumber\\
&&\hspace{-0mm}
+~\ln 2-4\gamma_E~+~O\big({z^+_{12}\over |z_{12_\perp}|\sigma}\big)\Big]e^{-ik^- z_1^+ +ik '^-z_2^+}
\label{Kexp}
\end{eqnarray}
one easily obtains
\begin{eqnarray}
&&\hspace{-1mm}
\calo^{\sigma_2}(z_{1+},z_{2+})~=~
e^{-2\balfa_s\ln{\sigma_2\over\sigma_1}[\ln\sigma_1\sigma_2+4\gamma_E-\ln 2\big]}
\nonumber\\
&&\hspace{-1mm}
\times~
\!\int\! dz'^+_1 dz'^+_2 ~\calo^{\sigma_1}(z'^+_1,z'^+_2)~z_{12_\perp}^{-2\balfa_s\ln{\sigma_2\over\sigma_1}}
\nonumber\\
&&\hspace{-1mm}
\times~{1\over 4\pi^2}\bigg[{i\Gamma\big(1-2\balfa_s\ln{\sigma_2\over\sigma_1}\big)
\over (z_1^+ - z'^+_1 + i\epsilon)^{1-2\balfa_s\ln{\sigma_2\over\sigma_1}}}
+c.c.\bigg]
\nonumber\\
&&\hspace{-1mm}
\times~
\bigg[{i\Gamma\big(1-2\balfa_s\ln{\sigma_2\over\sigma_1}\big)
\over (z_2^+ - z'^+_2 + i\epsilon)^{1-2\balfa_s\ln{\sigma_2\over\sigma_1}}}
+c.c.\bigg]
\label{result}
\end{eqnarray}
where we introduced notation $\balfa_s\equiv{\alpha_sN_c\over 4\pi}$.
It should be mentioned that the factor $4\gamma_E$ is ``scheme-dependent'': if one introduces to $\alpha$-integrals 
smooth cutoff $e^{-\alpha/a}$ instead of rigid cutoff $\theta(a>\alpha)$, the value $4\gamma_E$ changes
to $2\gamma_E$.

It is easy to see that the r.h.s. of Eq. (\ref{result}) transforms covariantly under all transformations 
(\ref{generators}) except Lorentz boost generated by $M^{+-}$. The reason is that the Lorentz boost
in $z$ direction changes cutoffs for the evolution. To understand 
that, note that  Eq. (\ref{Kexp}) is valid until $\sigma>{z^+_{12}\over z_{12_\perp}^2}$ 
so the  linear evolution (\ref{result}) is applicable in the region between
 \begin{equation}
\sigma_2=\sigma_B= {|z_{12_\perp}|\over z^-_{12}\sqrt{2}}~~~~~{\rm and}~~~~~\sigma_1={z^+_{12}\sqrt{2}\over |z_{12_\perp}|} 
 \label{fla1a}
 \end{equation}
 From Eq. (\ref{result}) it is easy to see that
 Lorentz boost  $z^+\!\rightarrow \!\lambda z^+, ~z^-\!\rightarrow\! {1\over\lambda} z^-$ changes the value of target
matrix element $\langle p_A|\calo|p_B\rangle$ by $\exp\{4\lambda\balfa_s\ln{z_{12\parallel}^2\over z_{12\perp}^2}\}$ but simultaneously it will change the 
result of similar evolution for projectile matrix element $\langle p_A|\ticalo|p_A\rangle$ by $\exp\{-4\lambda\balfa_s\ln{z_{12\parallel}^2\over z_{12\perp}^2}\}$ 
so the overall result for the amplitude (\ref{facoord}) remains intact.

\subsection{Evolution of quark TMD operators}

A simple calculation of evolution
of quark operator
\begin{eqnarray}
&&\hspace{-0mm}
\calo_q(z_1^+,z_{1_\perp};z_2^+,z_{2_\perp})~\equiv~
g^{2C_F\over b}\bsi(z_\perp+un) 
\\
&&\hspace{-0mm}
\times~[un+z_\perp,-\infty n]\not\!n
[z_\perp-\infty n,-\infty n][\infty n,0]\psi(0)
\nonumber
\end{eqnarray}
 the same evolution (\ref{result}) as for the gluon operators
 with trivial replacement $N_c\rightarrow C_F$
 \footnote{We assume that
 $\calo^q$ is defined with the same type of cutoff (\ref{cutoff}) imposed on quarks and gluons emitted 
as a result of evolution of $\calo^q$.}.
The factor $g^{2C_F\over b}$ 
($b\equiv {11\over 3}N_c-{2\over 3}n_f$) is added to avoid taking into account quark self-energy.

\subsection{Evolution beyond Sudakov region}

As we mentioned above, TMD factorization formula (\ref{TMDf}) for particle production 
at  $q_\perp\ll Q$ translates to the coordinate space  as Eq. (\ref{facoord}) 
with the requirement $z_{12_\parallel}^2\ll z_{12_\perp}^2$. As the result of evolution (\ref{result})
the transverse separation between gluon operators $\calf_i$ and $\calf_j$  remains intact while 
the longitudinal separation increases. As discussed in Refs. \cite{Balitsky:2015qba,Balitsky:2016dgz}  , the Sudakov approximation can be trusted until 
the upper cutoff in $\alpha$ integrals is greater than ${q_\perp^2\over x_Bs}$ which is equivalent to Eq. (\ref{sudapprox}) 
in the coordinate space.  If $x_B\sim 1$ and $q_\perp\sim m_N$, the relative energy between Wilson-line operators $\calf$
and target nucleon at the final point of evolution is  $\sim m_N^2$  
so one should use phenomenological models of TMDs with this low rapidity cutoff as a starting point of the evolution (\ref{result}).
If, however, $x_B\ll1$, this relative energy is  ${q_\perp^2\over x_B}\gg m_N^2$  
so one can continue the rapidity evolution in the region ${q_\perp^2\over x_Bs}>\sigma> {m_N^2\over s}$ 
beyond the Sudakov region into the small-$x$ region.  
The evolution in a ``proper'' small-$x$ region is known \cite{Dominguez:2011gc} - the TMD operator, known also as Weiczs\" acker-Williams
distribution,  will produce a hierarchy of color dipoles as a result of the non-linear evolution. 
However, the transition between Sudakov region and small-$x$ region is described by rather 
complicated interpolation formula \cite{Balitsky:2015qba}. In the coordinate space this means the study 
 of operator
$\calo$ at $z_\parallel^2\sim z_\perp^2$ and we hope that conformal considerations can help us
to obtain the TMD evolution in that region.

\section{Discussion}
As we mentioned in the Introduction, TMD evolution is analyzed by very different methods at small $x$ and moderate $x\sim 1$. 
In view of future EIC accelerator, which will probe the region between small $x$ and $x\sim 1$, we need a universal description 
of TMD evolution valid at both limits. Since the two formalisms differ even at the leading order where QCD 
is conformally invariant, our idea is to make this universal description first in ${\cal N}=4$ SYM.  As a first step, we found 
a conformally invariant evolution in the Sudakov region using our small-$x$ cutoff with the ``conformal refinement''  (\ref{cutoff}).

To compare with conventional TMD analysis let us write down the evolution of ``generalized TMD''\cite{Meissner:2009ww,Lorce:2013pza}
\begin{eqnarray}
&&\hspace{-0mm}
D^\sigma(x,\xi)~=~
\int\! dz^+e^{-ix\sqrt{s\over 2}z^+}\langle p'_B|\calo^\sigma\big(-{z^+\over 2},{z^+\over 2}\big)|p_B\rangle
\nonumber
\end{eqnarray}
where $\xi=-{p'_B-p_B\over \sqrt{2s}}$.
From Eq. (\ref{result}) one easily obtains
\begin{equation}
\hspace{-0mm}
{D^{\sigma_2}(x,\xi)\over D^{\sigma_1}(x,\xi)}~=~e^{-2\balfa_s\ln{\sigma_2\over\sigma_1}
[\ln\sigma_2\sigma_1(x^2-\xi^2)sz_{12_\perp}^2+4\gamma_E-2\ln 2]}
\label{momresult}
\end{equation}
For usual TMD at  $\xi=0$ with the limits of Sudakov evolution set by Eq. (\ref{fla1a}) one obtains
\begin{equation}
\hspace{-0mm}
{D^{\sigma_2}(x,q_\perp)\over D^{\sigma_1}(x,q_\perp)}
~=~e^{-2\balfa_s\ln{Q^2\over q_\perp^2}\big[\ln{Q^2\over q_\perp^2}+4\gamma_E-2\ln 2\big]}
\label{vperesult}
\end{equation}
which coincides with usual one-loop evolution of TMDs \cite{Aybat:2011zv} up to replacement $4\gamma_E-2\ln 2\rightarrow4\gamma_E-4\ln 2$.
As we discussed, such constant depends on the way of cutting $k^-$-integration which should be
coordinated with the cutoffs in the ``coefficient function'' $\sigma(ff\rightarrow H)$ in Eq. (\ref{TMDf}).
Thus,  the discrepancy is just like using two different schemes for usual renormalization. It should be mentioned, however,
that at $\xi\neq 0$ the result (\ref{momresult}) differs from conventional one-loop result which does not depend on $\xi$ , see e.g. \cite{Echevarria:2016mrc}.

Our main outlook is to try to connect to small-$x$ region, first in ${\cal N}=4$ and then in QCD.  Also,
it would be interesting to study if the ``conventional'' Sudakov-region results in two \cite{Echevarria:2015byo,Li:2016axz,Luebbert:2016itl}  
and three loops \cite{Echevarria:2016scs} 
can be recast in our cutoff  scheme which in principle allows transition to small-$x$ region. The study is in progress.

\begin{acknowledgments}
\section*{Acknowledgments}
\label{sec:acknowledgments}
We thank  V.M. Braun, A. Vladimirov and A. Tarasov for  discussions. 
The work of I.B.  was supported by DOE contract
 DE-AC05-06OR23177  and by the grant DE-FG02-97ER41028.

\end{acknowledgments}

\bibliography{tmdconfb.bib}

\end{document}